\documentstyle[amsfonts,twocolumn,graphicx,aps]{revtex}

\draft

\begin{document}
\title{Relativistic treatment of harmonics from impurity systems in quantum wires}
\author{O.A.~Castro-Alvaredo$^*$, A.~Fring$^*$, C.~Figueira de Morisson Faria$^{**}$}
\address{$*$Institut f\"{u}r Theoretische Physik, Freie Universit\"{a}t
Berlin, Arnimallee 14, D-14195 Berlin, Germany \\
$^{**}$Max Born Institut, Max Born Stra\ss e 2A, D-12489 Berlin, Germany \\
$^{**}$Institut f\"{u}r Theoretische Physik, Technische Universit\"at Wien,
Wiedner Hauptstra\ss e 8-10, A-1040 Wien, Austria}
\date{revised version: \today}
\maketitle

\begin{abstract}
Within a one particle approximation of the Dirac equation we investigate a
defect system in a quantum wire. We demonstrate that by minimally coupling a
laser field of frequency $\omega$ to such an impurity system, one may
generate harmonics of multiples of the incoming frequency. In a double
defect system one may use the distance between the defects in order to tune
the cut-off frequency.
\end{abstract}



\section{Introduction}

Since the early nineties, the generation of high-order harmonics of a
strong, low-frequency laser field has attracted a lot of attention in the
atomic physics community, see e.g. \cite{hhgreview,record} for a fairly
recent review. Indeed, high-harmonic generation has opened a wide range of
possibilities for obtaining high-frequency coherent sources, converting
infrared input radiation of frequency $\omega $ into light in the extreme
ultraviolet regime whose frequencies are multiples of $\omega $ (even up to
order $\sim 1000$, see e.g. \cite{record} for a recent review).

In gases, composed of atoms or small molecules, this phenomenon is
well-understood and, to some extent, even controllable in the sense that the
frequency of the highest harmonic, the so-called ``cut-off'', can be tuned
as well as the intensities of particular groups of harmonics. In more
complex systems, however, as for instance solids, or larger molecules,
high-harmonic generation is still an open problem. This is due to the fact
that, until a few years ago, such systems were expected not to survive the
strong laser fields involved. However, nowadays, with the advent of
ultrashort pulses, there exist solid-state materials whose damage threshold
is beyond the required intensities of $10^{14}{\rm W/cm}^{2}$ \cite{solid1}.
As a direct consequence, there is an increasing interest in such materials
as potential sources for high-harmonics. In fact, several groups are
currently investigating this phenomenon in systems such as thin crystals 
\cite{mois98,thincryst2}, carbon nanotubes \cite{mois2000}, or organic
molecules \cite{benz1,mois2001}.

Prototype solid-state devices are quantum wires, which nowadays do not only
serve as theoretical test ground, but may even be studied experimentally,
e.g. \cite{Mill}. With regard to the previously outlined problematic it is
of great interest to investigate how such devices interact with laser light.
In particular, the question of whether such systems are suitable
high-harmonic sources has not been dealt with up to now. This is the central
question which we shall be answering positively in this manuscript.

A useful particularity of quantum wires is that they are the physical
realization of models involving only one spatial dimension. Considering
theories in one space dimension has the further virtue that it allowed for
the development of various powerful non-perturbative techniques, which
exploit the integrability of the models. For instance, one may compute the
conductance of quantum wires in this fashion, see e.g. \cite{OA13} and
references therein. This also suggests that an exact relativistic treatment
of the electrons in the wire in the presence of the laser field is possible.
Therefore such a treatment constitutes an advantage over most approaches
adopted in the literature, in the context of atoms in strong laser fields,
which involve a series of approximations. The first, and most general
approximation performed in this context is that a proper field-theoretical
treatment is usually not taken into account. Furthermore, when dealing with
a relativistic situation, the Dirac equation is solved mostly by numerical
methods, which by themselves involve a series of approximation and exhibit a
good convergence only for high-frequency fields \cite{diraceq}. Further
approximations include the expansion \cite{expansions} of the Dirac equation
for the weakly relativistic case, the solution of the Klein-Gordon equation 
\cite{kleing,Grob} or of the relativistic Schr\"{o}dinger equation, or
purely classical treatments\cite{classical}. For some recent general reviews
see e.g. \cite{Review}.

In this paper, we solve the Dirac equation for an electron in a quantum wire
subject to an external laser field, including a single and multiple defects.
We investigate in various different regimes the radiation emitted by an
electron in this system in connection with the transmission at the defects
and demonstrate that the generation of harmonics is possible in such systems.

\section{Defect systems in laser fields}

Since the work of Weyl \cite{Weyl}, one knows that matter may be coupled to
light by means of a local gauge transformation, which reflects itself in the
usual minimal coupling prescription, i.e. $\partial _{\mu }\rightarrow
\partial _{\mu }-ieA_{\mu }$, with $A_{\mu }$ being the vector gauge
potential. For a free Fermion with mass $m$ this yields to the Lagrangian
density 
\begin{equation}
{\cal L}_{A}=\bar{\psi}(i\gamma ^{\mu }\partial _{\mu }-m+e\gamma ^{\mu
}A_{\mu })\psi \,.  \label{La}
\end{equation}
As common conventions\footnote{%
We adopt here relativistic units $1=c=\hbar =m\approx e^{2}137$ as mostly
used in the particle physics context rather than atomic units $1=e=\hbar
=m\approx c/137$ useful in atomic physics.}, we abbreviated here $\bar{\psi}%
=\psi ^{\dagger }\gamma ^{0}$ and use the following realization for the
gamma matrices 
\begin{equation}
\gamma ^{0}=\left( 
\begin{array}{cc}
0 & 1 \\ 
1 & 0
\end{array}
\right) ,\quad \quad \gamma ^{1}=\left( 
\begin{array}{cc}
0 & 1 \\ 
-1 & 0
\end{array}
\right) ,\quad \gamma ^{5}=\gamma ^{0}\gamma ^{1}.
\end{equation}
In the absence of the laser field the equations of motion for the free
Fermion may be solved by the well-known Fourier decomposition 
\begin{equation}
\psi _{j}^{f}(x,t)=\int \frac{d\theta }{\sqrt{4\pi }}(a_{j}(\theta
)u_{j}(\theta )e^{-i\vec{p}_{j}\cdot \vec{x}}+a_{\bar{\jmath}}^{\dagger
}(\theta )v_{j}(\theta )e^{i\vec{p}_{j}\cdot \vec{x}}).\qquad  \label{free}
\end{equation}
We parameterize the momentum as common through the rapidity $\theta $ by $%
p_{j}^{0}=m_{j}\cosh \theta $, $p_{j}^{1}=m_{j}\sinh \theta $ and denote the
anti-particle (positron) of the Fermion (electron) $j$ by $\bar{\jmath}$ .
For the Weyl spinors we employ the normalization 
\begin{equation}
u_{j}(\theta )=-i\gamma ^{5}v_{j}(\theta )=\sqrt{\frac{m_{j}}{2}}\left( 
\begin{array}{c}
e^{-\theta /2} \\ 
e^{\theta /2}
\end{array}
\right) \,,  \label{WS}
\end{equation}
and the creation and annihilation operators $a_{i}(\theta ),\,a_{i}^{\dagger
}(\theta )$ for a particle with rapidity $\theta $ obey the usual fermionic
anti-commutation relations $\{a_{i}(\theta _{1}),a_{j}(\theta _{2})\}=0$, $%
\{a_{i}(\theta _{1}),a_{j}^{\dagger }(\theta _{2})\}=2\pi \delta _{ij}\delta
(\theta _{1}-\theta _{2})$. When the laser field is switched on, we can
solve the equation of motion associated to (\ref{La}) 
\begin{equation}
(i\gamma ^{\mu }\partial _{\mu }-m+e\gamma ^{\mu }A_{\mu })\psi =0
\end{equation}
by a Gordon-Volkov type solution \cite{GV} 
\begin{eqnarray}
\psi _{j}^{A}(x,t) &=&\exp \left[ ie\int^{x}dsA_{1}(s,t)\right] \psi
_{j}^{f}(x,t),  \label{lo} \\
&=&\exp \left[ ie\int^{t}dsA_{0}(x,s)\right] \psi _{j}^{f}(x,t)\,.
\label{lo2}
\end{eqnarray}
Using now a linearly polarized laser field along the direction of the wire,
the vector potential can typically be taken in the dipole approximation to
be a superposition of monochromatic light with frequency $\omega $, i.e., 
\begin{eqnarray}
A(t) &:= &A_{1}(t)=\frac{1}{x}\int_{0}^{t}dsA_{0}(s)  \label{A01} \\
&=&-\frac{1}{2}\int_{0}^{t}dsE(s)=-\frac{E_{0}}{2}\int_{0}^{t}dsf(s)\cos
(\omega s)  \label{EF1}
\end{eqnarray}
with $f(t)$ being an arbitrary enveloping function equal to zero for $t<0$
and $t>\tau $, such that $\tau $ denotes the pulse length. In the following
we will always take $f(t)=\Theta (t)\Theta (\tau -t)\,$,$\ $with $\Theta (x)$
being the Heavyside unit step function. The second equality in (\ref{A01}), $%
A_{0}(x,t)=x\dot{A}(t)$, follows from the fact that we have to solve (\ref
{lo}) and (\ref{lo2}).

One comment is due with regard to the validity of the dipole approximation
in this context. It consists usually in neglecting the spatial dependence of
the laser field, which is justified when $x\omega <c=1$, where $x$ is a
representative scale of the problem considered. In the context of atomic
physics this is typically the Bohr radius. In the problem investigated here,
this approximation has to hold over the full spatial range in which the
Fermion follows the electric field. We can estimate this classically, in
which case the maximal amplitude is $eE_{0}/\omega ^{2}$ and therefore the
following constraint has to hold 
\begin{equation}
\left( \frac{eE_{0}}{\omega }\right) ^{2}=4U_{p}<1\,,  \label{Dipole}
\end{equation}
for the dipole approximation to be valid. Due to the fact that $x$ is a
function of $\omega $, we have now a lower bound on the frequency rather
than an upper one as is more common in the context of atomic physics. We
have also introduced here the ponderomotive energy $U_{p}$ for monochromatic
light, that is the average kinetic energy transferred from the laser field
to the electron.

The solutions to the equations of motion of the free system and the one
which includes the laser field are then related by a factor similar to the
gauge transformation from the length to the velocity gauge 
\begin{equation}
\psi _{j}^{A}(x,t)=\exp \left[ ixeA(t)\right] \psi _{j}^{f}(x)\,.
\label{LLa}
\end{equation}
In an analogous fashion one may use the same minimal coupling procedure also
to couple in addition the laser field to the defect. One has to invoke the
equation of motion in order to carry this out, since as in \cite{OA13}, we
assume here also that the defect is linear in the fields $\bar{\psi}$ and $%
\psi $. The Lagrangian density for a complex free Fermion $\psi $ with $\ell 
$ defects ${\cal D}^{\alpha }(\bar{\psi},\psi ,A_{\mu })$ of type $\alpha $
at the position $x_{n}$ subjected to a laser field then reads 
\begin{equation}
{\cal L}_{AD}={\cal L}_{A}+\sum\limits_{n=1}^{\ell }{\cal D}^{\alpha _{n}}(%
\bar{\psi},\psi ,A_{\mu })\,\delta (x_{n})\,.  \label{Lda}
\end{equation}
Considering for simplicity first the case of a single defect situated at $%
x=0 $, the solution to the equation of motion resulting from (\ref{Lda}) is
taken to be of the form 
\begin{equation}
\psi _{j}^{A}(x,t)=\Theta (x)\psi _{j,+}^{A}(x,t)+\Theta (-x)\psi
_{j,-}^{A}(x,t)\,.  \label{An}
\end{equation}
This means we distinguish here by notation the solutions (\ref{LLa}) on the
left and right of the defect, $\psi _{j,-}^{A}(x,t)$ and $\psi
_{j,+}^{A}(x,t)$, respectively. This is also reflected in the corresponding
creation and annihilation operators $a_{j,+}(\theta )$, $a_{j,-}(\theta )$,
etc. One may then proceed according to standard potential scattering theory
and notes that these functions are not independent of each other.
Substitution of (\ref{An}) into the equation of motion yields the
constraints 
\begin{equation}
i\gamma ^{1}(\psi _{j,+}^{A}(x,t)-\psi _{j,-}^{A}(x,t))|_{x=0}=\left. \frac{%
\partial {\cal D}_{AD}(\bar{\psi},\psi ,A_{\mu })}{\partial \bar{\psi}%
_{j}^{A}(x,t)}\right| _{x=0}.  \label{bcon}
\end{equation}
These restrictions (\ref{bcon}) serve to determine the transmission and
reflection amplitudes. The extension to multiple defects, that is having
equations of the type (\ref{bcon}) for each defect situated at position $%
x=x_{n}$, follows in an analogous straightforward manner.

\subsection{Transmission and Reflection amplitudes}

Substituting the Fourier decomposition (\ref{free}), together with the free
Fermion solution minimally coupled to a laser field (\ref{LLa}) into the
constraint (\ref{bcon}), one can determine the reflection and transmission
amplitudes from the left to the right, $R$ and $T$, respectively. They are
defined in an obvious manner as 
\begin{equation}
a_{j,-}(\theta )=R_{_{j}}(\theta )a_{j,-}(-\theta )+T_{_{j}}(\theta
)a_{j,+}(\theta )\,.
\end{equation}
When parity invariance is broken, the corresponding amplitudes from the
right to the left do not have to be identical and are defined as 
\begin{equation}
a_{j,+}(-\theta )=\tilde{T}_{_{j}}(\theta )a_{j,-}(-\theta )+\tilde{R}%
_{j}(\theta )a_{j,+}(\theta )\,.
\end{equation}
In this way the laser field parameters $E_{0}$ and $\omega $ will be
quantities on which the $R$'s and $T$'s depend upon at a particular time $t$%
. When iterating these equations one obtains the corresponding expressions
for multiple defect systems, e.g. placing for instance the defect ${\cal D}%
^{\alpha }$ left from ${\cal D}^{\beta }$ one obtains the well known
expressions 
\begin{eqnarray}
T_{i}^{\alpha \beta }(\theta ) &=&\frac{T_{i}^{\alpha }(\theta )T_{i}^{\beta
}(\theta )}{1-R_{i}^{\beta }(\theta )\tilde{R}_{i}^{\alpha }(\theta )},
\label{TD} \\
R_{i}^{\alpha \beta }(\theta ) &=&R_{i}^{\alpha }(\theta )+\frac{%
R_{i}^{\beta }(\theta )T_{i}^{\alpha }(\theta )\tilde{T}_{i}^{\alpha
}(\theta )}{1-R_{i}^{\beta }(\theta )\tilde{R}_{i}^{\alpha }(\theta )}\,.
\label{RD}
\end{eqnarray}

\noindent From our previous comment on the validity of the dipole
approximation it is clear that the distance between the two defects
introduces a new scale in the system, which has to be constraint as $%
y<\omega ^{-1}$. In addition, to justify that the multiple defect system can
be treated effectively as a single one, requires that the sum of the $y$'s
is much smaller than the length of the wire. Similar expressions, which we
will not need in what follows, hold for the parity reversed situation and
for more defects, see e.g. \cite{OA13} and references therein.

In addition with regard to the application of high harmonic generation, we
shall be interested in the spectrum of frequencies which are filtered out by
the defect while the laser pulse is non-zero. The Fourier transforms of the
reflection and transmission probabilities provide exactly this information 
\begin{eqnarray}
{\cal T}(\Omega ,\theta ,E_{0},\omega ,\tau ) &=&\frac{1}{\tau }%
\int_{0}^{\tau }dt|T(\theta ,E_{0},\omega ,t)|^{2}\cos (\Omega t),
\label{TT} \\
{\cal R}(\Omega ,\theta ,E_{0},\omega ,\tau ) &=&\frac{1}{\tau }%
\int_{0}^{\tau }dt|R(\theta ,E_{0},\omega ,t)|^{2}\cos (\Omega t).
\end{eqnarray}
When parity is preserved for the reflection amplitudes, we have $%
|T|^{2}+|R|^{2}=1$, and it suffices to consider ${\cal T}$ \ in the
following.

\subsubsection{Type I defects}

Taking the laser field in form of monochromatic light in the dipole
approximation (\ref{EF1}), we may naturally assume that the transmission
probability for some particular defects can be expanded as 
\begin{equation}
|T_{I}(\theta ,U_{p},\omega ,t)|^{2}=\sum_{k=0}^{\infty }t_{2k}(\theta
)(4U_{p})^{k}\sin ^{2k}(\omega t).  \label{exp}
\end{equation}
We shall refer to defects which admit such an expansion as ``type I
defects''. Assuming that the coefficients $t_{2k}(\theta )$ become at most $%
1 $, we have to restrict our attention to the regime $4U_{p}<1$ in order for
this expansion to be meaningful for all $t$. Note that this is no further
limitation, since it is precisely the same constraint as already encountered
for the validity of the dipole approximation (\ref{Dipole}). The first
coefficient is always the transmission probability for vanishing laser
field, that is $t_{0}(\theta )=|T(\theta ,E_{0}=0)|^{2}$. The functional
dependence of (\ref{exp}) will turn out to hold for various explicit defects
considered below. Based on this equation, we compute for such type of defect 
\begin{equation}
{\cal T}_{I}(\Omega ,\theta ,U_{p},\omega ,\tau )=\sum_{k=0}^{\infty }\frac{%
(2k)!(U_{p})^{k}\sin (\tau \Omega )t_{2k}(\theta )}{\tau \Omega
\prod_{l=1}^{k}[l^{2}-(\Omega /2\omega )^{2}]}\,.  \label{ts}
\end{equation}
It is clear from this expression that type I defects will preferably let
even multiples of the basic frequency $\omega $ pass, whose amplitudes will
depend on the coefficients $t_{2k}(\theta )$. When we choose the pulse
length to be integer cycles, i.e. $\tau =2\pi n/\omega =nT$ for $n\in {\Bbb Z%
}$, the expression in (\ref{ts}) reduces even further. The values at even
multiples of the basic frequency are simply 
\begin{equation}
{\cal T}_{I}(2n\omega ,\theta ,U_{p})=(-1)^{n}\sum_{k=0}^{\infty
}t_{2k}(\theta )\left( U_{p}\right) ^{k}\left( 
\begin{array}{c}
2k \\ 
k-n
\end{array}
\right) ,  \label{tss}
\end{equation}
which becomes independent of the pulse length $\tau $. Notice also that the
dependence on $E_{0}$ and $\omega $ occurs in the combination of the
ponderomotive energy $U_{p}$. Further statements require the precise form of
the coefficients $t_{2k}(\theta )$ and can only be made with regard to a
more concrete form of the defect.

\subsubsection{Type II defects}

Clearly, not all defects are of the form (\ref{exp}) and we have to consider
also expansions of the type 
\begin{equation}
|T_{II}(\theta ,E_{0}/e,\omega ,t)|^{2}=\sum_{k,p=0}^{\infty
}t_{2k}^{p}(\theta )\frac{E_{0}^{2k+p}}{\omega ^{2k}}\cos ^{p}(\omega t)\sin
^{2k}(\omega t).  \label{exp2}
\end{equation}
We shall refer to defects which admit such an expansion as ``type II
defects''. In this case we obtain 
\begin{eqnarray}
&&{\cal T}_{II}(\Omega ,\theta ,E_{0}/e,\omega ,\tau )=\sum_{k,p=0}^{\infty
}\sum_{l=0}^{p}\left( 
\begin{array}{c}
p \\ 
l
\end{array}
\right) \frac{\Omega \sin (\tau \Omega )}{(-1)^{l+1}\tau \omega ^{2+2k}} \\
&&\left( \frac{(2k+2l)!t_{2k}^{2p}(\theta )}{\prod%
\limits_{q=0}^{k+l}[(2q)^{2}-(\frac{\Omega }{\omega })^{2}]}+\frac{%
(2k+2l)!t_{2k}^{2p+1}(\theta )E_{0}}{\prod\limits_{q=1}^{k+l+1}[(2q-1)^{2}-(%
\frac{\Omega }{\omega })^{2}]}\right) E_{0}^{2k+2p}.  \nonumber
\end{eqnarray}
We observe from this expression that type II defects will filter out all
multiples of $\omega $. For\ the pulse being once again of integer cycle
length, this reduces to 
\begin{eqnarray}
{\cal T}_{II}(2n\omega ,\theta ,U_{p},E_{0}) &=&\sum_{k,p=0}^{\infty
}\sum_{l=0}^{p}(-1)^{l+n}\frac{t_{2k}^{2p}(\theta )}{2^{2l-2p}}\left(
U_{p}\right) ^{k+p}  \nonumber \\
&&\times E_{0}^{2p}\left( 
\begin{array}{c}
p \\ 
l
\end{array}
\right) \left( 
\begin{array}{c}
2k+2l \\ 
k+l-n
\end{array}
\right)  \label{II1}
\end{eqnarray}
and 
\begin{eqnarray}
&&{\cal T}_{II}((2n-1)\omega ,\theta ,E_{0}/e)=\sum_{k,p=0}^{\infty
}\sum_{l=0}^{p}(-1)^{l+n+1}\frac{t_{2k}^{2p+1}(\theta )}{2^{2l-2p+1}} 
\nonumber \\
&&\times \left( U_{p}\right) ^{k+p}\left( 
\begin{array}{c}
p \\ 
l
\end{array}
\right) \frac{(2k+2l)!(2n-1)E_{0}^{2p+1}}{(l+k-n+1)!(l+n+k)!}  \label{II2}
\end{eqnarray}
which are again independent of $\tau $. We observe that in this case we can
not combine the $E_{0}$ and $\omega $ into a $U_{p}$. The analytical
expressions presented in this section will not only serve as a benchmark for
our analytical computation, but can be used in addition to extract various
structural information as we see below.

\subsection{One particle approximation}

In spite of the fact that we are dealing with a quantum field theory, it is
known that a one particle approximation to the Dirac equation is very useful
and physically sensible when the external forces vary only slowly on a scale
of a few Compton wavelengths, see e.g. \cite{IZ}. We may therefore define
the spinor wavefunctions 
\begin{eqnarray}
\Psi _{j,u,\theta }(x,t) &:&=\psi _{j}^{A}(x,t)\frac{\left| a_{j}^{\dagger
}(\theta )\right\rangle }{\sqrt{2\pi ^{2}p_{j}^{0}}}=\frac{e^{-i\vec{p}%
_{j}\cdot \vec{x}}}{\sqrt{2\pi p_{j}^{0}}}u_{j}(\theta ) \\
\Psi _{j,v,\theta }(x,t)^{\dagger } &:&=\psi _{j}^{A}(x,t)^{\dagger }\frac{%
\left| a_{j}^{\dagger }(\theta )\right\rangle }{\sqrt{2\pi ^{2}p_{j}^{0}}}=%
\frac{e^{-i\vec{p}_{j}\cdot \vec{x}}}{\sqrt{2\pi p_{j}^{0}}}v_{j}(\theta
)^{\dagger }\,.
\end{eqnarray}
With the help of these functions we obtain then for the defect system 
\begin{eqnarray}
&&{\bf \Psi }_{i,u,\theta }^{A}(x,t):=\psi _{i}^{A}(x,t)\frac{\left|
a_{i,-}^{\dagger }(\theta )\right\rangle }{\sqrt{2\pi ^{2}p_{i}^{0}}}= 
\nonumber \\
&&\Theta (-x)\left[ \Psi _{i,u,\theta }(x,t)+\Psi _{i,u,-\theta
}(x,t)R_{i}^{\ast }(\theta )\right] +  \nonumber \\
&&\Theta (x)T_{_{i}}^{\ast }(\theta )\left[ \Psi _{i,u,\theta }(x,t)+\Psi
_{i,u,-\theta }(x,t)\tilde{R}_{_{i}}^{\ast }(-\theta )\right]  \label{fx}
\end{eqnarray}
and the same function with $u\rightarrow v$. Since this function resembles a
free wave it can not be normalized properly and we have to localize the wave
in form of a wave packet by multiplying with an additional function, $\tilde{%
g}(p,t)$ in (\ref{free}) and its counterpart $g(x,t)$ in (\ref{fx}),
typically a Gau\ss ian. Then for the function ${\bf \Phi }_{i,u,\theta
}^{A}(x,t)=g(x,t){\bf \Psi }_{i,u,\theta }^{A}(x,t)$, we can achieve that $%
\left\| {\bf \Phi }\right\| =1$.

\section{Harmonic generation}

As mentioned above, harmonic generation has attracted large attention in the
atomic physics community in recent years. It is the non-linear response of a
dipole moment, in general atomic, to an external laser field. Here we want
to investigate whether such responses also exist for defect systems. We
carry out our treatment relativistically. The time dependent dipole moment
is given by 
\begin{equation}
x_{j,u,\theta }(t)=\left\langle {\bf \Phi }_{j,u,\theta }^{A}(x,t)^{\dagger
}x{\bf \Phi }_{j,u,\theta }^{A}(x,t)\right\rangle  \label{xt}
\end{equation}
such that the emission spectrum is the absolute value of the Fourier
transform of the dipole moment 
\begin{equation}
{\cal X}_{j,u,\theta }(\Omega )=\left| \int_{0}^{\tau }dt\,x_{j,u,\theta
}(t)\exp i\Omega t\right| \,\,.
\end{equation}
We localize now the wave packet in a region much smaller than the classical
estimate for the maximal amplitude the electron will acquire when following
the laser field. We achieve this with a Gau\ss ian $g(x,t)=\exp
(-x^{2}/\Delta )$, where $\Delta \ll eE_{0}/\omega ^{2}$. Placing the defect
at the origin and neglecting its extension, the computation of (\ref{xt})
with (\ref{fx}) then boils down to the evaluation of 
\begin{eqnarray}
&&x_{i,u,\theta }(t)=\int\limits_{-\infty }^{0}\frac{x\left[ 1+|R_{i}(\theta
)|^{2}+2%
\mathop{\rm Re}%
\frac{e^{2ix\sinh \theta }R_{i}(\theta )}{\cosh \theta }\right] }{2\pi e^{%
\frac{2}{\Delta }[x+eA(t)]^{2}}}dx+  \nonumber \\
&&\int\limits_{0}^{\infty }\frac{x|T_{i}(\theta )|^{2}\left[ 1+|R_{i}(\theta
)|^{2}+2%
\mathop{\rm Re}%
\frac{e^{2ix\sinh \theta }R_{i}(\theta )^{\ast }}{\cosh \theta }\right] }{%
2\pi e^{\frac{2}{\Delta }[x+eA(t)]^{2}}}dx\,.  \label{33}
\end{eqnarray}
The expressions for the $R$'s and $T$'s depend of course on the form of the
defect and further generic statements can not be made at this point. We
therefore turn to a concrete example.

\section{The energy operator defect}

Localizing sharply the energy operator $\varepsilon (x)=g\bar{\psi}\psi (x)$%
, with $g$ being a coupling constant, yields a defect which has been studied
extensively in the absence of a laser field. It should be noted that this is
only a particular example and one may also consider other type of defects in
a analogous fashion, see \cite{OA13} and references therein. One of the
virtues of this defect is that it is real, thus preserving parity invariance.

Coupling the vector potential minimally to this type of defect yields 
\begin{equation}
{\cal D}_{AD}(\bar{\psi},\psi ,A_{\mu })=g\bar{\psi}(1+e/m\gamma ^{\mu
}A_{\mu })\psi \,,
\end{equation}
by invoking the equation of motion.

\subsection{Transmission and Reflection amplitudes}

We are now interested in determining the reflection and transmission
amplitudes associated to this defect by the potential scattering method as
outlined in section II A. Taking from now on $m=1$, we compute for the
various reflection amplitudes 
\begin{eqnarray}
&&R_{i}(\theta ,g,A/e,y)=\tilde{R}_{i}(\theta ,g,-A/e,-y)=  \nonumber \\
&&R_{\bar{\imath}}(\theta ,g,A/e,-y)=\tilde{R}_{\bar{\imath}}(\theta
,g,-A/e,y)=  \label{r1} \\
&&\frac{[y\dot{A}-\cosh \theta ]e^{-2iy\sinh \theta }}{[1-y\dot{A}\cosh
\theta ]-i\frac{g}{4}[\frac{4}{g^{2}}+1+A^{2}-y^{2}\dot{A}^{2}]\sinh \theta }%
\,.  \nonumber
\end{eqnarray}
We denoted the differentiation with respect to time by a dot. The
transmission amplitudes turn out to be 
\begin{eqnarray}
&&T_{i}(\theta ,g,A/e,y)=\tilde{T}_{i}(\theta ,g,-A/e,-y)=  \nonumber \\
&&T_{\bar{\imath}}(\theta ,g,-A/e,y)=\tilde{T}_{\bar{\imath}}(\theta
,g,A/e,-y)=  \label{t1} \\
&&\frac{i\left[ 1-y^{2}\dot{A}^{2}+(A-\frac{2i}{g})^{2}\right] \sinh \theta 
}{\frac{4}{g}[1-y\dot{A}\cosh \theta ]-i[\frac{4}{g^{2}}+1+A^{2}-y^{2}\dot{A}%
^{2}]\sinh \theta }\,.  \nonumber
\end{eqnarray}
Locating the defect at $\ y=0$, the derivative of $A$ does not appear
anymore explicitly in (\ref{r1}) and (\ref{t1}), such that it is clear that
this defect is of type I and admits an expansion of the form (\ref{exp}).
With the explicit expressions (\ref{r1}) and (\ref{t1}) at hand, we can
determine all the coefficients $t_{2k}(\theta )$ in (\ref{exp})
analytically. For this purpose let us first bring the transmission amplitude
into the more symmetric form 
\begin{equation}
\left| T_{i}(\theta ,g,A/e)\right| ^{2}=\frac{\tilde{a}_{0}(\theta
,g)+a_{2}(\theta ,g)A^{2}+a_{4}(\theta ,g)A^{4}}{a_{0}(\theta
,g)+a_{2}(\theta ,g)A^{2}+a_{4}(\theta ,g)A^{4}},  \label{tt}
\end{equation}
with 
\begin{eqnarray}
a_{0}(\theta ,g) &=&16g^{2}+(4+g^{2})^{2}\sinh ^{2}\theta , \\
\tilde{a}_{0}(\theta ,g) &=&(g^{2}-4)^{2}\sinh ^{2}\theta ,\qquad \\
\,\,a_{2}(\theta ,g) &=&2g^{2}(4+g^{2})\sinh ^{2}\theta , \\
a_{4}(\theta ,g) &=&g^{4}\sinh ^{2}\theta .\qquad
\end{eqnarray}
We can now expand $\left| T(\theta ,g,A)\right| ^{2}$ in powers of the field 
$A(t)$ and identify the coefficients $t_{2k}(\theta ,g)$ in (\ref{exp})
thereafter. To achieve this we simply have to carry out the series expansion
of the denominator in (\ref{tt}). The latter admits the following compact
form 
\[
\frac{1}{a_{0}(\theta ,g)+a_{2}(\theta ,g)A^{2}+a_{4}(\theta ,g)A^{4}}%
=\sum_{k=0}^{\infty }c_{2k}(\theta ,g)A^{2k}, 
\]
with $c_{0}(\theta ,g)=1/a_{0}(\theta ,g)$ and 
\[
c_{2k}(\theta ,g)=-\frac{c_{2k-2}(\theta ,g)a_{2}(\theta ,g)+c_{2k-4}(\theta
,g)a_{4}(\theta ,g)}{a_{0}(\theta ,g)}, 
\]
for $k>0$. We understand here that all coefficients $c_{2k}$ with $k<0$ are
vanishing, such that from this formula all the coefficients $c_{2k}$ may be
computed recursively. Hence, by comparing with the series expansion (\ref
{exp}), we find the following closed formula for the coefficients $%
t_{2k}(\theta ,g)$%
\begin{equation}
t_{2k}(\theta ,g)=[\tilde{a}_{0}(\theta ,g)-a_{0}(\theta ,g)]c_{2k}(\theta
,g)\quad k>0.
\end{equation}
The first coefficients then simply read 
\begin{eqnarray}
t_{0}(\theta ,g) &=&\frac{\tilde{a}_{0}(\theta ,g)}{a_{0}(\theta ,g)}%
=|T(\theta ,E_{0}=0)|^{2}, \\
t_{2}(\theta ,g) &=&\frac{a_{2}(\theta ,g)}{a_{0}(\theta ,g)}\left[
1-t_{0}(\theta ,g)\right] \\
&=&\frac{8g^{4}(4+g^{2})\sinh ^{2}2\theta }{(16g^{2}+(4+g^{2})^{2}\sinh
^{2}\theta )^{2}}, \\
t_{4}(\theta ,g) &=&\left[ \frac{a_{4}(\theta ,g)}{a_{2}(\theta ,g)}-\frac{%
a_{2}(\theta ,g)}{a_{0}(\theta ,g)}\right] t_{2}(\theta ,g),
\end{eqnarray}
and so on. It is now clear how to obtain also the higher terms analytically,
but since they are rather cumbersome we do not report them here.

\subsection{Harmonic generation}

With the coefficients $t_{2k}$, we can compute the series (\ref{ts}) and (%
\ref{tss}) in principle to any desired order. For some concrete values of
the laser and defect parameters the results of our evaluation are depicted
in figure 1.

The main observation is that the defect acts as a filter selecting higher
harmonics of even order of the laser frequency. Furthermore, from the zoom
of the peak regions, we see that there are satellite peaks appearing near
the main harmonics. They reduce their intensity when $\tau $ is increased,
such that with longer pulse length the harmonics become more and more
pronounced.

We also investigated that for different frequencies $\omega $ the general
structure will not change. Increasing the field amplitude $E_{0}$, simply
lifts up the whole plot without altering very much its overall structure. We
support these findings in two alternative ways, either by computing directly
(\ref{TT}) numerically or, more instructively, by evaluating the sums (\ref
{ts}) and (\ref{tss}).
\smallskip
\includegraphics[width=8.2cm,height=6.09cm]{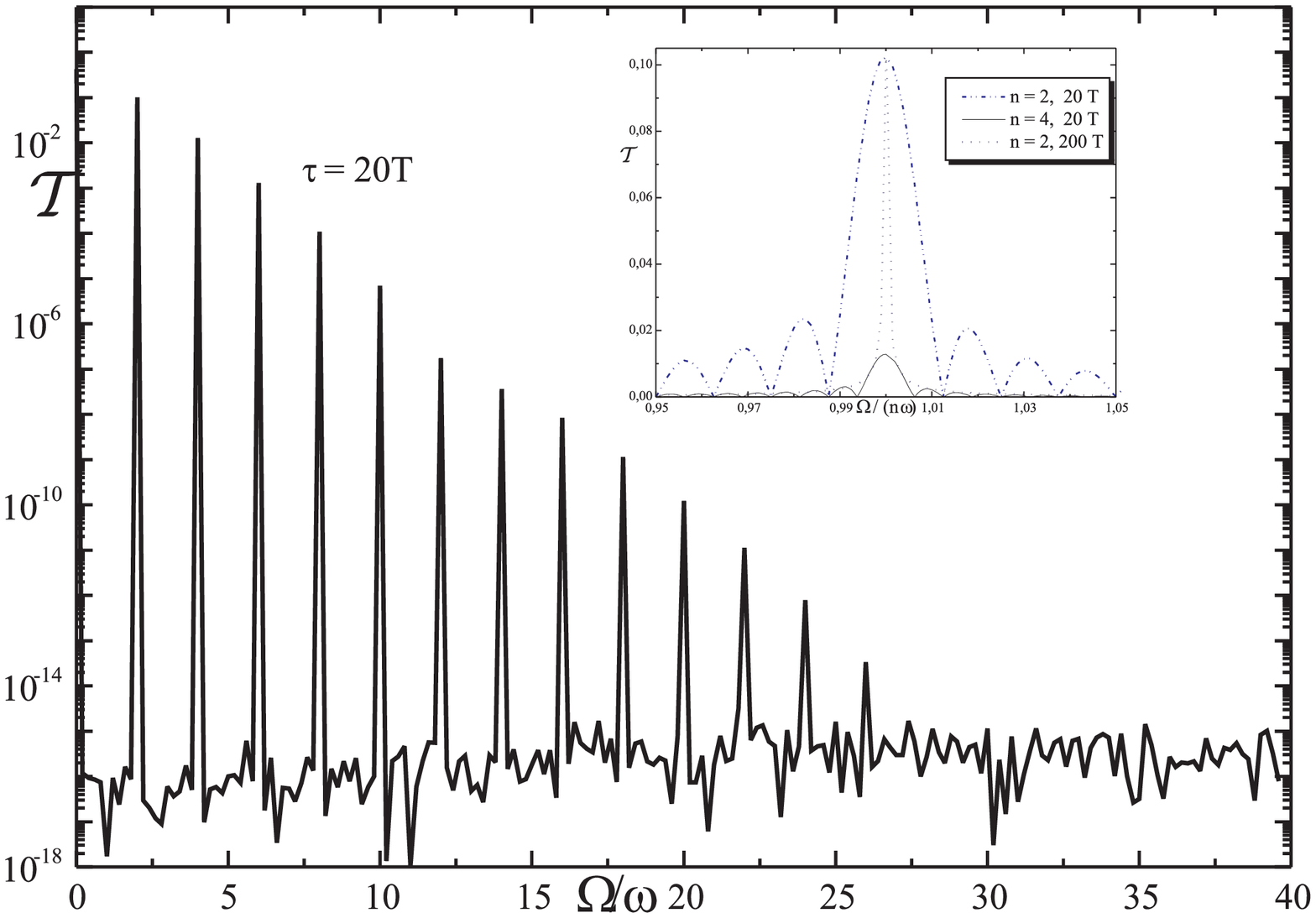}

\noindent {\small Figure 1: Absolute value squared
of the Fourier transform of the transmission
probability for a single defect with $E_{0}=2.0$, $g=3.5$, $\theta =1.2$, $%
\omega =0.2$. }
\medskip

Let us now carry out a similar analysis for a double defect system. We place
one of the defects at $x=0$ and the other at $x=y$. The distance emerges now
as a new scale in the system and note from our comment on the validity of
the dipole approximation that is restricted as $y<\omega ^{-1}$. In addition 
$y$ has to be much smaller than the total length of the wire. The double
defect amplitudes are computed directly from (\ref{TD}) and (\ref{RD}) with
the expression for the single defect (\ref{r1}) and (\ref{t1}). Since now
both $A$ and $\dot{A}$ appear explicitly in the formulae for $R$'s and $T$%
's, it is clear that the expansion of the double defect can not be of type
I, but it turns out to be of type II, i.e., of the form (\ref{exp2}). Hence,
we will now expect that besides the even also the odd multiples of $\omega $
will be filtered out. This is confirmed by our explicit computations for two
identical defects as depicted in figure 2.

\smallskip
\includegraphics[width=8.2cm,height=6.09cm]{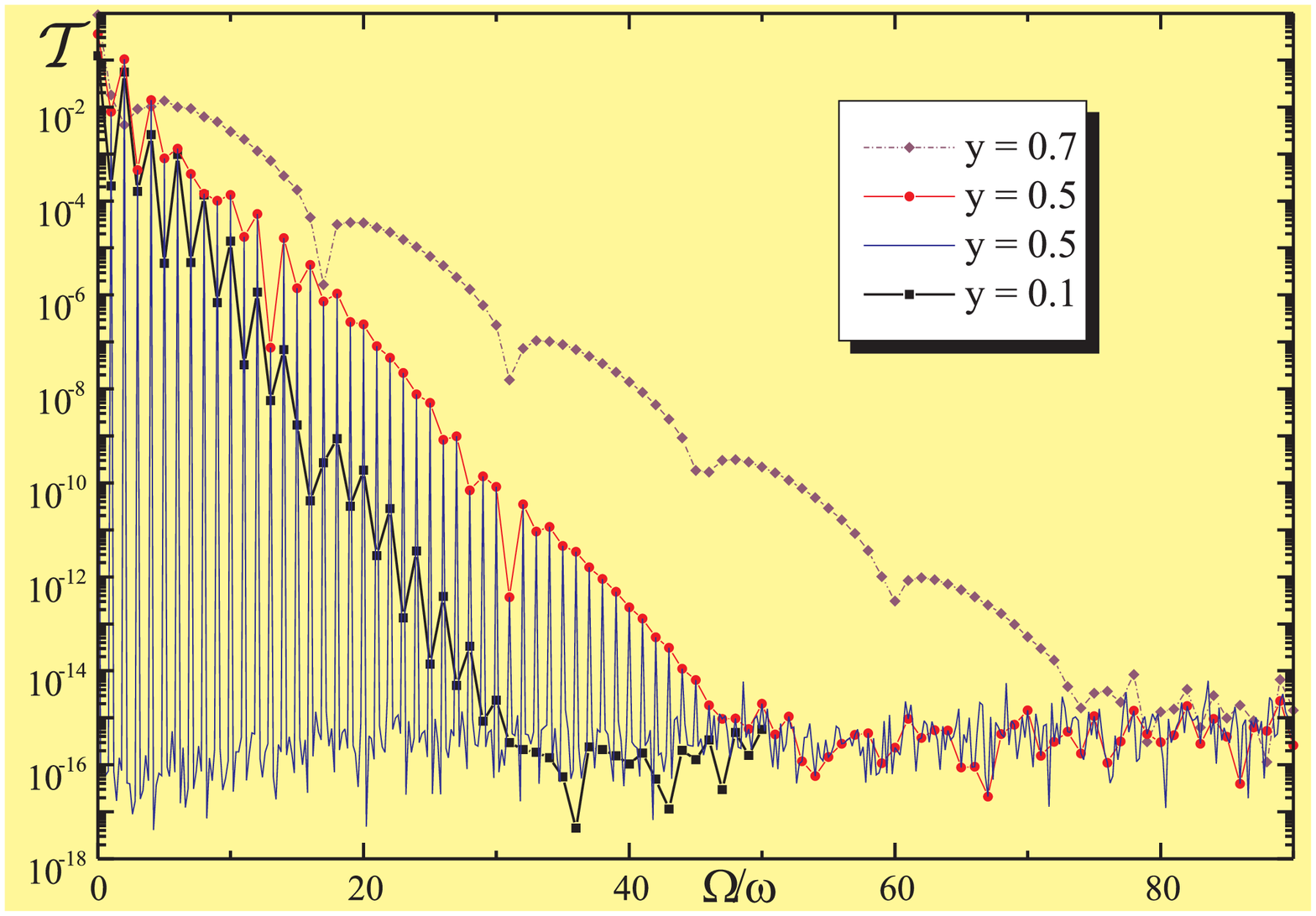}

\noindent {\small Figure 2: Absolute value squared
of the  Fourier transform of the transmission
probability for a double defect with $E_{0}=2.0$, $g=3.5$, $\theta =1.2$, $%
\omega =0.2$ and varying $y$.}
\medskip

Here we have only plotted a continuous spectrum for $y=0.5$, whereas for
reasons of clarity, we only drew the enveloping function which connects the
maxima of the harmonics for the remaining distances. We observe that now not
only odd multiples of the frequency emerge in addition as harmonics, but
also that we obtain much higher harmonics and the cut-off is shifted further
to the ultraviolet. Furthermore, we see a periodic pattern in the enveloping
function, which appears to be independent of $y$. Similar patterns were
observed before in the literature, as for instance in the context of atomic
physics described by a Klein-Gordon formalism (see figure 2 in \cite{Grob}).

\smallskip

\includegraphics[width=8.2cm,height=6.09cm]{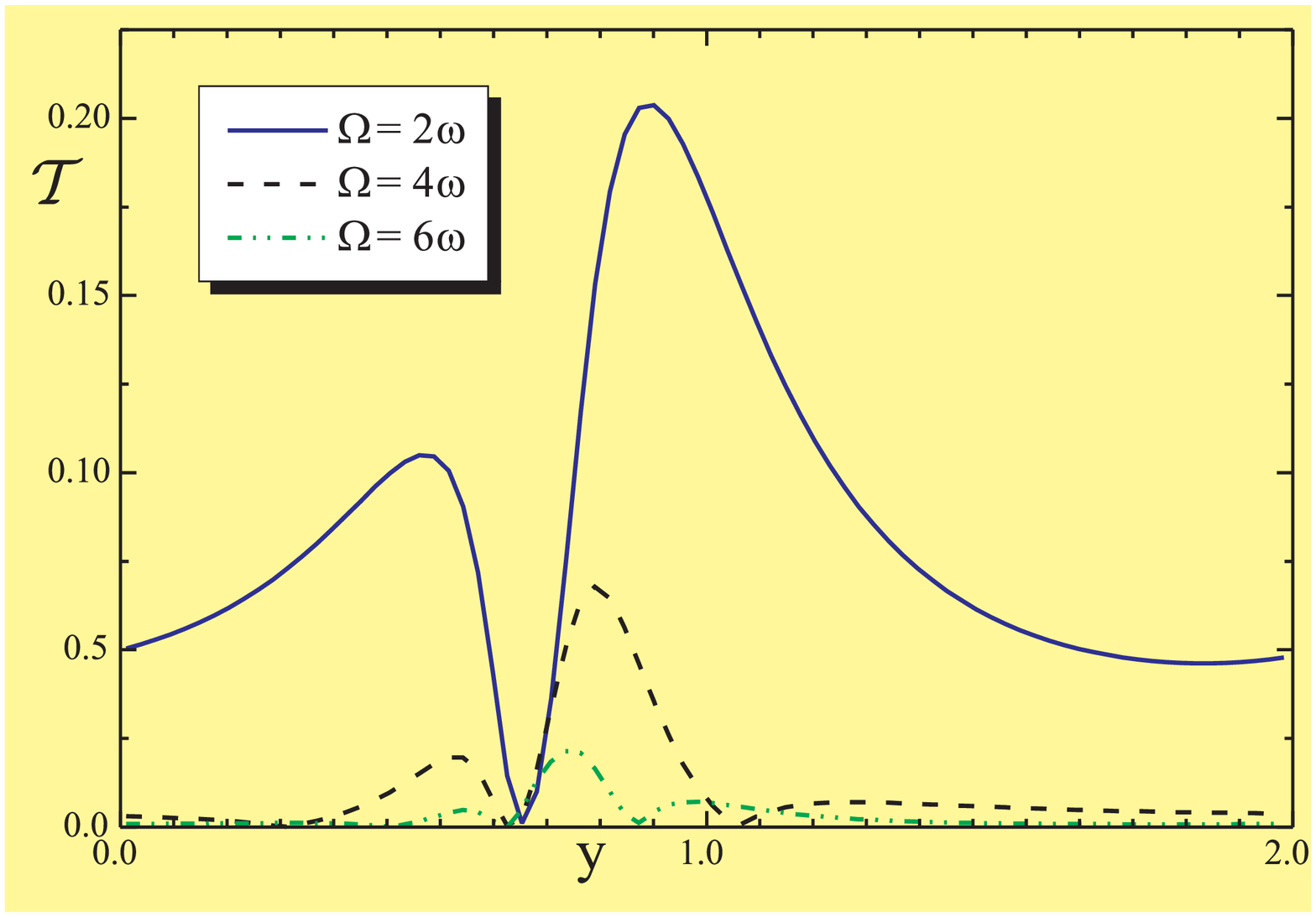}

\includegraphics[width=8.2cm,height=6.09cm]{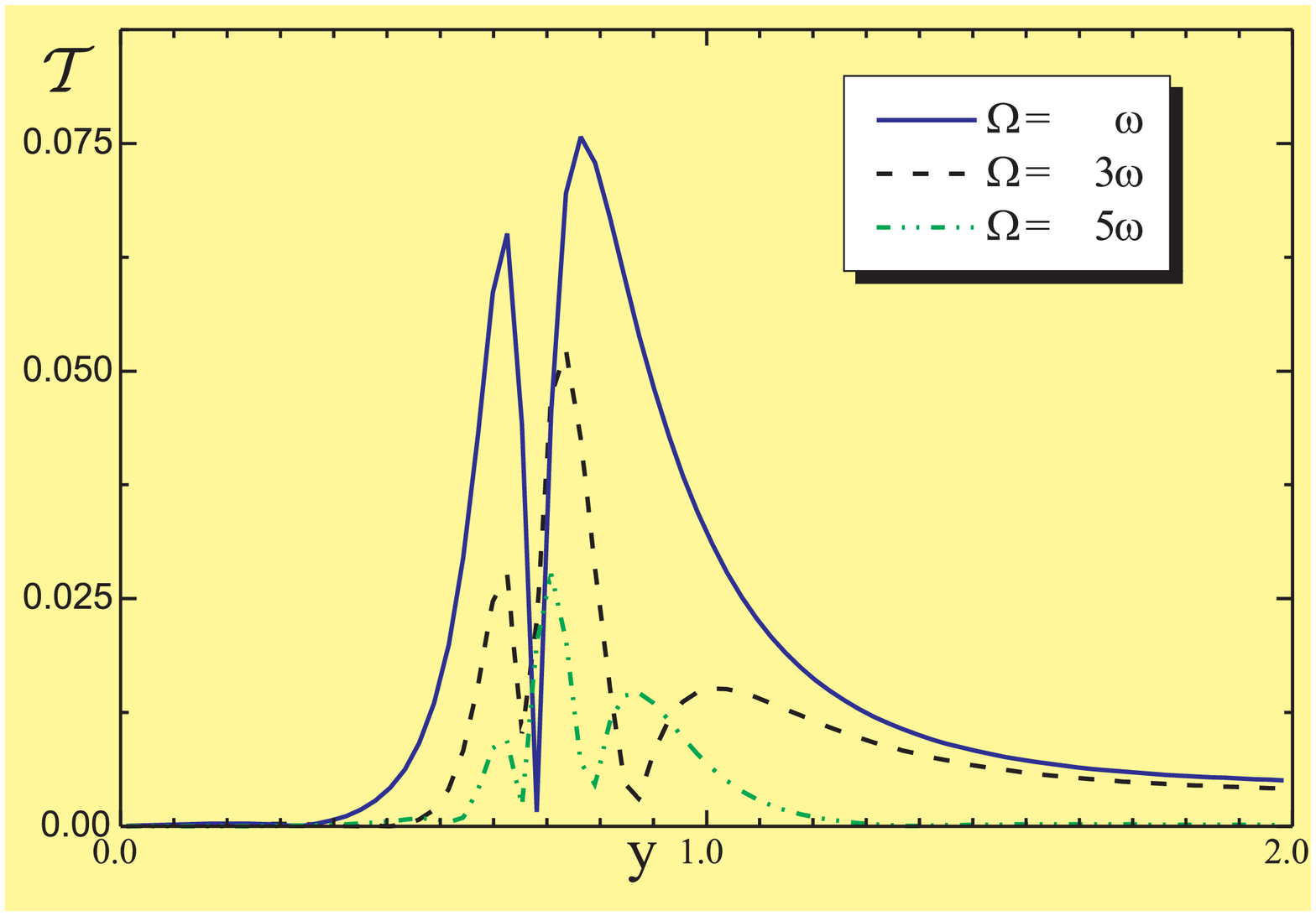}

\noindent {\small Figure 3: Absolute value squared
of the Fourier transform of the transmission
probability for a double defect with $E_{0}=2.0$, $g=3.5$, $\theta =1.2$,
for even and odd multiples of $\omega =0.2$ and varying $y$. } \medskip

With varying distance $y$ the new structures can be modulated, i.e. we can
control the intensity of certain peaks and also shift the cut-off. Clearly
for a concrete application one would like the control mechanism to be as
simple as possible and therefore it is interesting to investigate precisely
how the emission amplitude and the cut-off behave as functions of $y$.
Unfortunately, this function turns out to be not very simple as can be seen
from figure 3, where we present our analysis for varying $y$ and particular
fixed harmonics. Nonetheless, there is a universal shape common to all
harmonics of the same type. As expected from the analytical expressions the
overall pattern for the odd and even harmonics differs. We confirm our
previous observation, namely that for small values of the distance $y$,
which corresponds to the limit of a single defect, there are no odd
harmonics emerging.

Let us now come to the main point of our analysis and see how this structure
is reflected in the harmonic spectrum. The result of the evaluation of (\ref
{33}) is depicted in figure 4.\smallskip

\includegraphics[width=8.2cm,height=6.09cm]{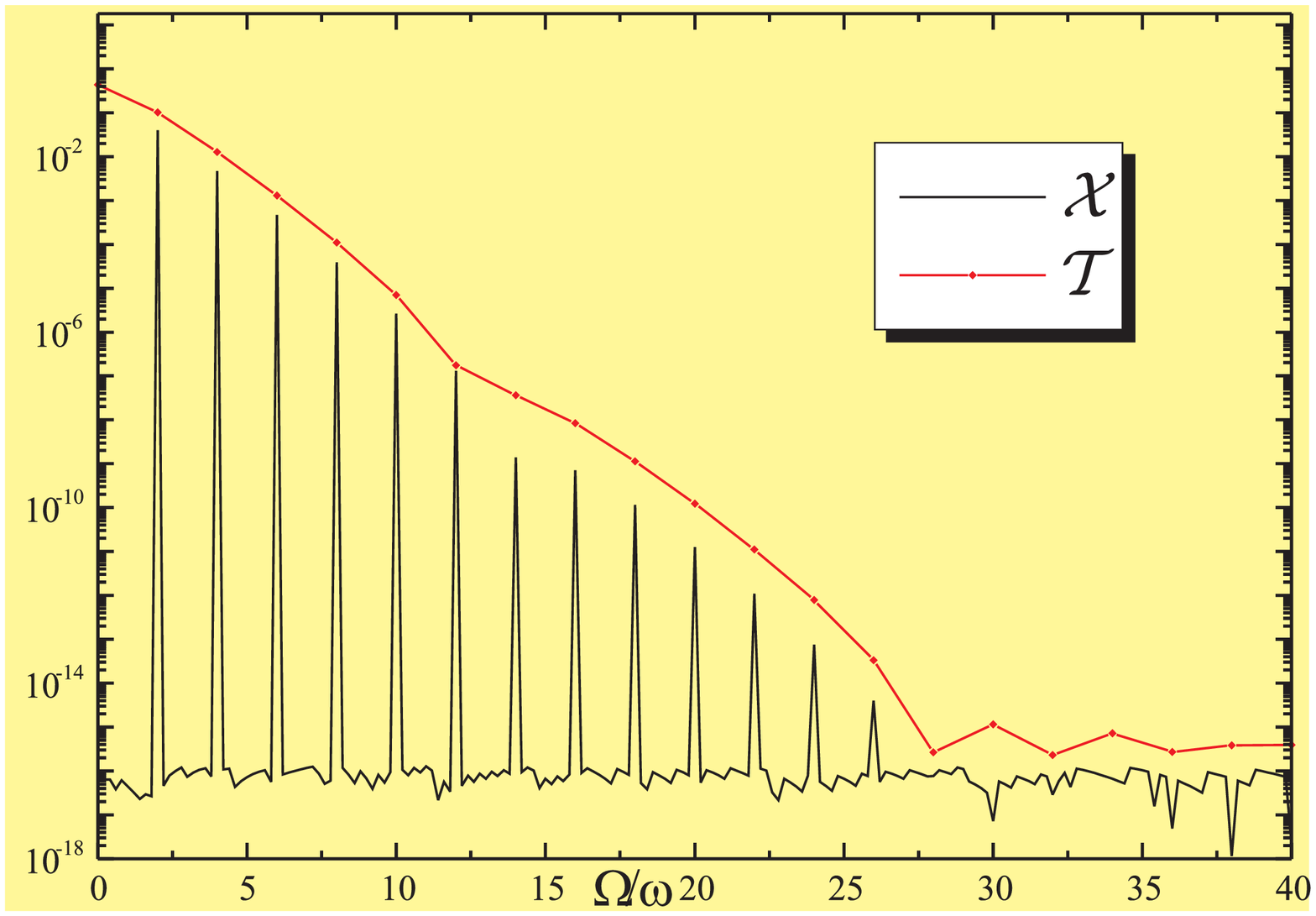}

\noindent {\small Figure 4: Harmonic emission spectrum for a single defect
with $E_{0}=2.0$, $g=3.5$, $\theta =1.2$, $\omega =0.2$, $\Delta =6$. }
\medskip

We observe a very similar spectrum as we have already computed for the
Fourier transform of the transmission amplitude, which is not entirely
surprising with regard to the expression (\ref{33}). The cut-off frequencies
are essentially identical. From the comparison between ${\cal X}$ and the
enveloping function for ${\cal T}$ \ we deduce, that the term involving the
transmission amplitude clearly dominates the spectrum.

Let us now turn to the computation of the emission spectrum for a double
defect system coupled to a laser field. We depict the results of our
analysis in figure 5.

\includegraphics[width=8.2cm,height=6.09cm]{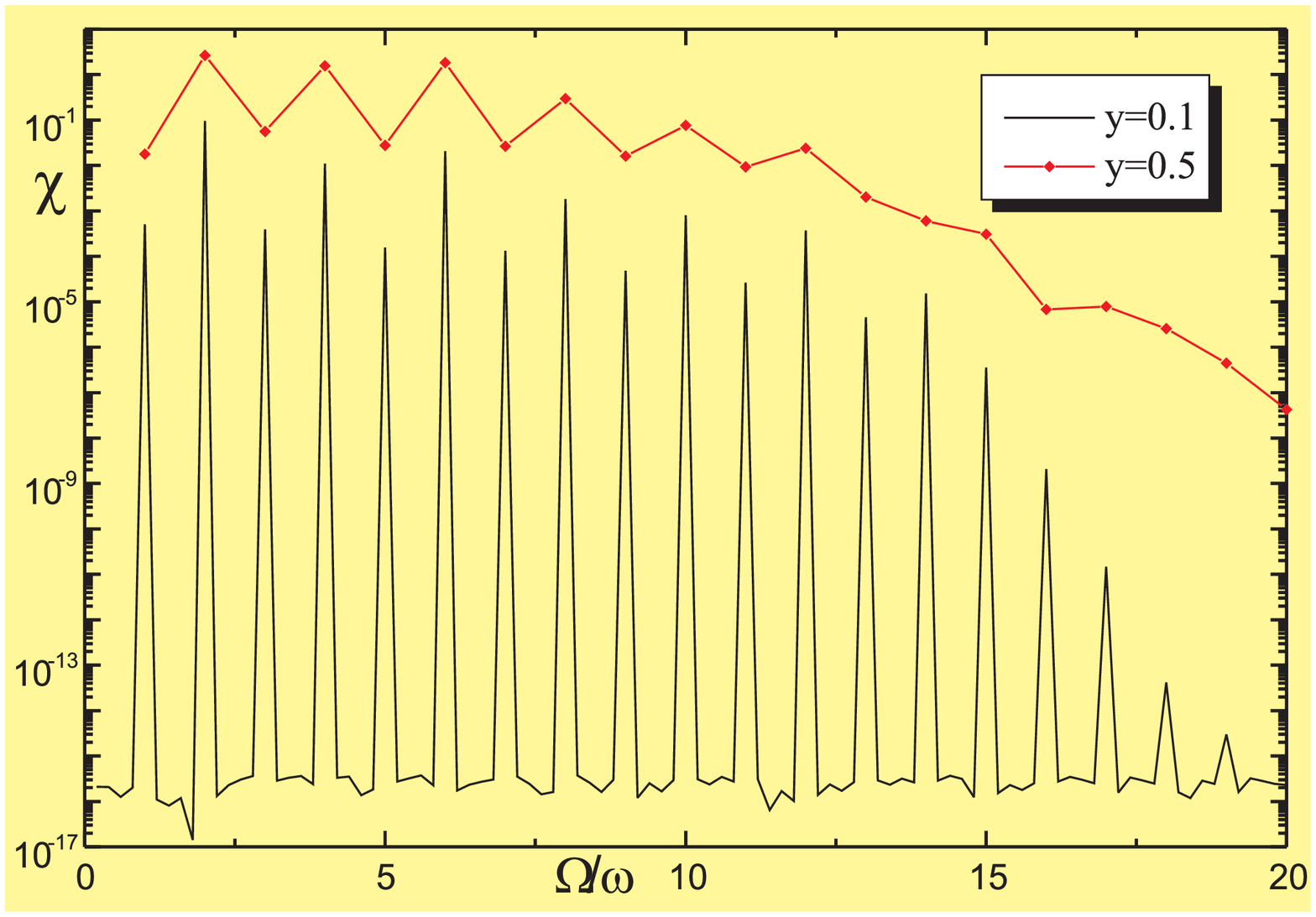}

\noindent {\small Figure 5: Harmonic emission spectrum for a double defect
with $E_{0}=2.0$, $g=3.5$, $\theta =1.2$, $\omega =0.2$, $\Delta =6$%
.\medskip }

Once again, we observe a qualitatively similar behaviour as for ${\cal T}$,
in particular the occurrence of even and odd order harmonics. We remark that
in our approach for larger values of $y$ the normalization of the wave
function becomes somewhat inaccurate and therefore the relative height in
the intensities is not very precise. In principle this could be compensated,
but for the conclusions we are trying to reach here that is not important.

The important general deduction from these computation is of course that
harmonics of higher order do emerge in the emission spectrum of a defect
system.

\subsection{Relativistic versus non-relativistic regime}

In the previous sections we have been working in an intensity regime which
is close to the damage threshold of a solid, according to the experimental
observations in \cite{solid1}. This allowed us to see the maximum effect
with regard to harmonic generation which at present might be visible from
experiments. However, it is also interesting to investigate situations which
are not experimentally feasible at present and of course lower intensity
regimes.

In order to judge in which regime we are working and whether there are
relativistic effects, let us carry out various limits. First of all we
recall a standard estimation according to which the relativistic kinetic
energy is close to the classical one when one is dealing with velocities $%
v^{2}\ll 3/4c^{2}$. This is the same as saying that the kinetic energy is
much smaller than the rest mass $E_{\text{kin}}\ll m_{0}c^{2}$. Making now a
rough estimation for the system under consideration, we assume that the
total kinetic energy is the one obtained from the laser field, i.e. the
ponderomotive energy $U_{p}$. We also ignore for this estimation any
sophisticated corrections, such as possible Doppler shifts in the frequency,
etc. Then the non-relativistic regime is characterized by the condition $%
U_{p}\ll 1$.

\includegraphics[width=8.2cm,height=6.09cm]{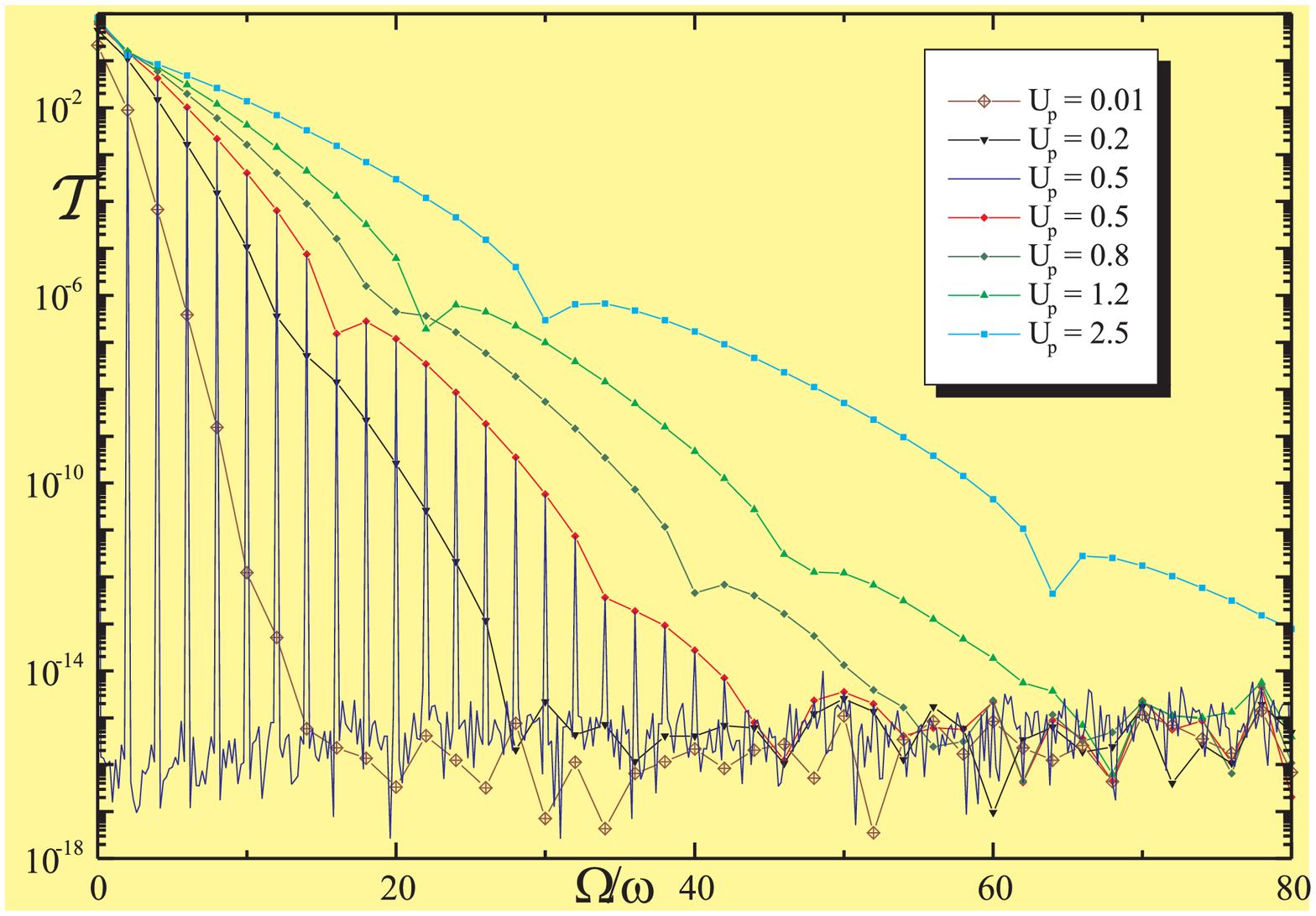}

\noindent {\small Figure 6: Absolute value squared
of the Fourier transform of the transmission
probability for a single defect for various values of \ }$U_{p}$ {\small %
with $g=3.5$, $\theta =1.2.$} \medskip

Based on our previous observation that ${\cal T}$ \ and ${\cal X}$ exhibit a
very similar behaviour, it will be sufficient here to study only the ${\cal T%
}$ \ in the different regimes, which will be easier than an investigation of
the full emission spectrum (\ref{33}). From our analytic expression (\ref
{tss}), we see that for a type I defect the quantity ${\cal T}_{I}$ becomes
a function of $U_{p}$, such that the regime will be the same when we rescale
simultaneously $E_{0}$ and $\omega $. Accordingly we evaluate numerically
the Fourier transform (\ref{TT}), or equivalently compare against our
analytical expression (\ref{tss}), and depict our results in figure 6.

We observe that when passing more and more towards the relativistic regime
the cut-off is increased. The other feature one recognizes is that the
modulating structure in the enveloping function of the harmonics becomes
more pronounced. One should also note, in regard to (\ref{Dipole}), that the
multipole structures might become more and more important in the
relativistic regime.

Let us now perform a similar computation for the double defect. From the
expressions (\ref{II1}) and (\ref{II2}) we see that now ${\cal T}_{II}$ is
not just a simple function of $U_{p}$ and therefore even being in the same
regime the behaviour will be different when $E_{0}$ and $\omega $ are
rescaled. We alter in that case the regimes by rescaling $E_{0}$ and keeping
the frequency fixed. Our results are depicted in figure 7.

\includegraphics[width=8.2cm,height=6.09cm]{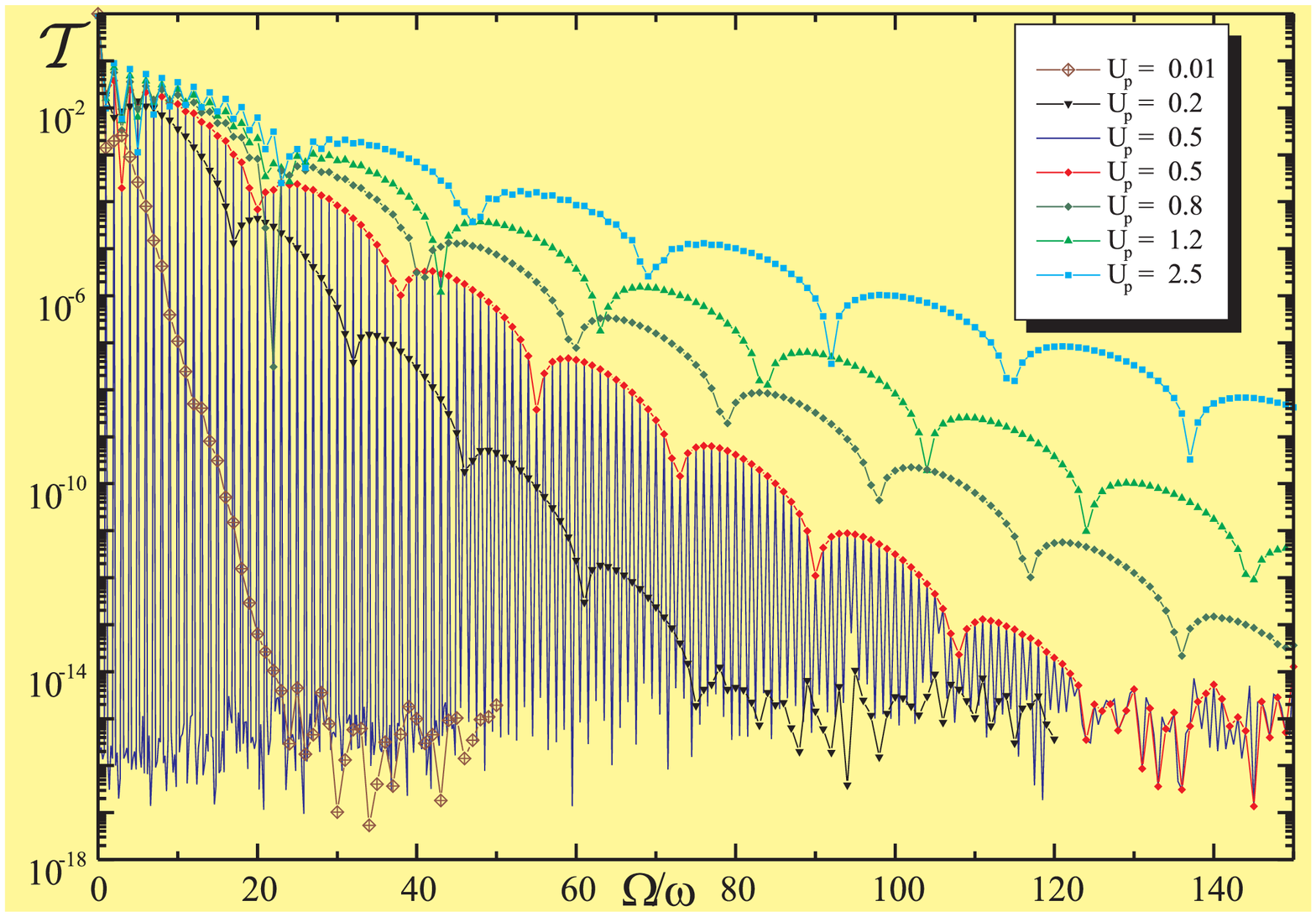} \medskip

\noindent {\small Figure 7: Absolute value squared
of the Fourier transform of the transmission
probability for a double defect for varying values of \ }$E_{0}$ {\small %
with $g=3.5$, $\theta =1.2,\omega =0.2$.} \medskip

Similar as for the single defect we see that the cut-off is increased and
the modulating structure in the enveloping function becomes more emphasized
when we move towards the relativistic regime. In addition we note that the
difference between the even and odd harmonic becomes larger with increasing $%
U_{p}$. This effect is more extreme for the low order harmonics.

As a general observation we state that there are not any effects which seem
to be special to the relativistic regime, but the transition to that regime
seems to be rather smooth.

\section{Conclusions}

We have minimally coupled a laser field to an impurity system in a quantum
wire described with the Dirac equation. Using a free field expansion for the
free Dirac Fermion, we computed by means of standard potential scattering
theory the reflection and transmission amplitudes associated to such type of
defect. The amplitudes become functions of the defect coupling constant $g$,
their separation $y$ and the laser field parameters $E_{0},\omega $ and $t$.
We employed these amplitudes in order to evaluate the emission spectrum of a
dipole moment in such a system. Our findings for a single defect, taken to
be the energy operator coupled minimally to the laser field, are that even
multiples of the driving frequency $\omega $ are emitted. Investigating a
double defect system of two of such defects, we observe the emission of odd
as well as even multiples of the original frequency. These features may
already be observed qualitatively on the Fourier expansion of the
transmission amplitude, even analytically. When carrying out the
non-relativistic and extreme relativistic limit we do not observe any
special effect, the transition seems to be rather smooth.

There are various questions left for further investigation. As an
interesting application one may for instance compute the conductance in a
similar fashion as in \cite{OA13} and employ the laser to control it.

\medskip \medskip

{\bf Acknowledgments:} O.A. C-A. and A.F. are grateful to the Deutsche
Forschungsgemeinschaft (Sfb288) for financial support. We thank E. Lenz for
providing various references.

\end{document}